\documentclass[reprint,superscriptaddress,amsmath,amssymb,aps,prc,twocolumn]{revtex4-2}

\usepackage{graphicx}
\usepackage{hyperref}
\usepackage{color}
\usepackage[english]{babel}
\usepackage{bm}
\usepackage[T1]{fontenc}
\usepackage{threeparttable}
\usepackage{multirow}

\usepackage{lineno,hyperref}

\usepackage{makecell}
\usepackage{array}

\begin{document}

\title{High-precision measurements of low-lying isomeric states in $^{120-124}$In\\
with JYFLTRAP double Penning trap}

\author{D.A.~Nesterenko}
\thanks{These authors contributed equally}
\affiliation{University of Jyvaskyla, Department of Physics, Accelerator laboratory, P.O. Box 35(YFL) FI-40014 University of Jyvaskyla, Finland}
\author{J.~Ruotsalainen}
\thanks{These authors contributed equally}
\email{jouni.k.a.ruotsalainen@jyu.fi}
\affiliation{University of Jyvaskyla, Department of Physics, Accelerator laboratory, P.O. Box 35(YFL) FI-40014 University of Jyvaskyla, Finland}
\author{M.~Stryjczyk}
\thanks{These authors contributed equally}
\email{marek.m.stryjczyk@jyu.fi}
\affiliation{University of Jyvaskyla, Department of Physics, Accelerator laboratory, P.O. Box 35(YFL) FI-40014 University of Jyvaskyla, Finland}
\author{A.~Kankainen}
\email{anu.kankainen@jyu.fi}
\affiliation{University of Jyvaskyla, Department of Physics, Accelerator laboratory, P.O. Box 35(YFL) FI-40014 University of Jyvaskyla, Finland}
\author{L.~Al~Ayoubi}
\affiliation{University of Jyvaskyla, Department of Physics, Accelerator laboratory, P.O. Box 35(YFL) FI-40014 University of Jyvaskyla, Finland}
\affiliation{Universit\'e Paris Saclay, CNRS/IN2P3, IJCLab, 91405 Orsay, France}
\author{O.~Beliuskina}
\affiliation{University of Jyvaskyla, Department of Physics, Accelerator laboratory, P.O. Box 35(YFL) FI-40014 University of Jyvaskyla, Finland}
\author{P.~Delahaye}
\affiliation{GANIL, CEA/DSM-CNRS/IN2P3, Boulevard Henri Becquerel, 14000 Caen, France}
\author{T.~Eronen}
\affiliation{University of Jyvaskyla, Department of Physics, Accelerator laboratory, P.O. Box 35(YFL) FI-40014 University of Jyvaskyla, Finland}
\author{M.~Flayol}
\affiliation{Universit\'e de Bordeaux, CNRS/IN2P3, LP2I Bordeaux, UMR 5797, F-33170 Gradignan, France}
\author{Z.~Ge}
\affiliation{University of Jyvaskyla, Department of Physics, Accelerator laboratory, P.O. Box 35(YFL) FI-40014 University of Jyvaskyla, Finland}
\affiliation{GSI Helmholtzzentrum f\"ur Schwerionenforschung, 64291 Darmstadt, Germany}
\author{W.~Gins}
\affiliation{University of Jyvaskyla, Department of Physics, Accelerator laboratory, P.O. Box 35(YFL) FI-40014 University of Jyvaskyla, Finland}
\author{M.~Hukkanen}
\affiliation{University of Jyvaskyla, Department of Physics, Accelerator laboratory, P.O. Box 35(YFL) FI-40014 University of Jyvaskyla, Finland}
\affiliation{Universit\'e de Bordeaux, CNRS/IN2P3, LP2I Bordeaux, UMR 5797, F-33170 Gradignan, France}
\author{A.~Jaries}
\affiliation{University of Jyvaskyla, Department of Physics, Accelerator laboratory, P.O. Box 35(YFL) FI-40014 University of Jyvaskyla, Finland}
\author{D.~Kahl}
\affiliation{Extreme Light Infrastructure – Nuclear Physics, Horia Hulubei National Institute for R\&D in Physics and Nuclear Engineering (IFIN-HH), 077125 Bucharest-M\u{a}gurele, Romania}
\author{D.~Kumar}
\affiliation{GSI Helmholtzzentrum f\"ur Schwerionenforschung, 64291 Darmstadt, Germany}
\author{S.~Nikas}
\affiliation{University of Jyvaskyla, Department of Physics, Accelerator laboratory, P.O. Box 35(YFL) FI-40014 University of Jyvaskyla, Finland}
\author{A. Ortiz-Cortes}
\affiliation{University of Jyvaskyla, Department of Physics, Accelerator laboratory, P.O. Box 35(YFL) FI-40014 University of Jyvaskyla, Finland}
\affiliation{GANIL, CEA/DSM-CNRS/IN2P3, Boulevard Henri Becquerel, 14000 Caen, France}
\author{H.~Penttil\"a}
\affiliation{University of Jyvaskyla, Department of Physics, Accelerator laboratory, P.O. Box 35(YFL) FI-40014 University of Jyvaskyla, Finland}
\author{D.~Pitman-Weymouth}
\affiliation{Department of Physics and Astronomy, University of Manchester, Manchester M13 9PL, United Kingdom}
\author{A.~Raggio}
\affiliation{University of Jyvaskyla, Department of Physics, Accelerator laboratory, P.O. Box 35(YFL) FI-40014 University of Jyvaskyla, Finland}
\author{M.~Ramalho}
\affiliation{University of Jyvaskyla, Department of Physics, Accelerator laboratory, P.O. Box 35(YFL) FI-40014 University of Jyvaskyla, Finland}
\author{M.~Reponen}
\affiliation{University of Jyvaskyla, Department of Physics, Accelerator laboratory, P.O. Box 35(YFL) FI-40014 University of Jyvaskyla, Finland}
\author{S.~Rinta-Antila}
\affiliation{University of Jyvaskyla, Department of Physics, Accelerator laboratory, P.O. Box 35(YFL) FI-40014 University of Jyvaskyla, Finland}
\author{J.~Romero}
\affiliation{University of Jyvaskyla, Department of Physics, Accelerator laboratory, P.O. Box 35(YFL) FI-40014 University of Jyvaskyla, Finland}
\affiliation{Department of Physics, Oliver Lodge Laboratory, University of Liverpool, Liverpool, L69 7ZE, United Kingdom}
\author{A.~de Roubin}
\altaffiliation[Present address: ]{KU Leuven, Instituut voor Kern- en Stralingsfysica, B-3001 Leuven, Belgium}
\affiliation{Universit\'e de Bordeaux, CNRS/IN2P3, LP2I Bordeaux, UMR 5797, F-33170 Gradignan, France}
\author{P.C.~Srivastava}
\affiliation{Department of Physics, Indian Institute of Technology Roorkee, Roorkee 247667, India}
\author{J.~Suhonen}
\affiliation{University of Jyvaskyla, Department of Physics, Accelerator laboratory, P.O. Box 35(YFL) FI-40014 University of Jyvaskyla, Finland}
\author{V.~Virtanen}
\affiliation{University of Jyvaskyla, Department of Physics, Accelerator laboratory, P.O. Box 35(YFL) FI-40014 University of Jyvaskyla, Finland}
\author{A.~Zadvornaya}
\altaffiliation[Present address: ]{II. Physikalisches Institut, Justus Liebig Universit\"at Gie{\ss}en, 35392 Gie{\ss}en, Germany}
\affiliation{University of Jyvaskyla, Department of Physics, Accelerator laboratory, P.O. Box 35(YFL) FI-40014 University of Jyvaskyla, Finland}

\date{\today}

\begin{abstract}
Neutron-rich $^{120-124}$In isotopes have been studied utilizing the double Penning trap mass spectrometer JYFLTRAP at the IGISOL facility. Using the phase-imaging ion-cyclotron-resonance technique, the isomeric states were resolved from ground states and their excitation energies measured with high precision in $^{121,123,124}$In. In $^{120,122}$In, the $1^+$ states were separated and their masses were measured while the energy difference between the unresolved $5^+$ and $8^-$ states, whose presence was confirmed by post-trap decay spectroscopy was determined to be $\leq15$~keV. In addition, the half-life of $^{122}$Cd, $T_{1/2} = 5.98(10)$~s, was extracted. Experimental results were compared with energy density functionals, density functional theory and shell-model calculations.

\end{abstract}

\maketitle

\section{Introduction}

Neutron-rich indium isotopes around the doubly-magic $^{132}$Sn nucleus are a center of attention and they are extensively studied via a plethora of experimental methods, such as mass measurements \cite{Kankainen2013,Babcock2018,Nesterenko2020,Izzo2021}, laser spectroscopy, \cite{Vernon2019,Vernon2022} decay spectroscopy \cite{Taprogge2014,Lorusso2015,Taprogge2016,Jungclaus2016,Piersa2018,Piersa2019,Phong2019,Dunlop2019,Benito2020,Whitmore2020,PiersaSilkowska2021,Phong2022} and transfer reactions \cite{Vaquero2020}. It is because nuclei around the doubly-magic core are relatively simple systems and they constitute a perfect testing ground for various theoretical approaches \cite{Yuan2016,Koura2017,Utama2018,Shimizu2021,HKWang2021,HKWang2022}. However, indium isotopes lying closer to the valley of stability remain largely unknown as the experimental data is very scarce \cite{ENSDF}. The issue is exacerbated for the odd-odd isotopes where only a few excited levels are known and the order of the long-lived states is not established at all \cite{NUBASE2020,ENSDF}.  

Particular isomeric states have been considered of astrophysical importance for many years \cite{Ward1980}. In recent years, more and more isomeric states are being added into astrophysical calculations \cite{Fujimoto2020,Misch2021,Wendell2021,Wendell2021b}. Due to differences in ground- and isomeric-state half-lives, the release of decay heat might be accelerated or delayed \cite{Fujimoto2020,Wendell2021b}. A sensitivity study on the isomers behavior indicated that unknown $\gamma$-ray transitions in the $^{120-124}$In isotopes have a non-negligible influence on the transition rates between the ground and the isomeric states \cite{Wendell2021}. These transitions are modelled using Weisskopf estimates; however, they rely on excitation energies which are not measured.

The masses of $^{120-124}$In are known with a relatively low precision as they were only measured with transfer reaction and $\beta$ end-point studies \cite{Wang2021}. These types of measurements are known to have systematic issues and deviate significantly from the more precise and accurate Penning-trap values \cite{Nesterenko2019,Ge2021,Eronen2022,Ramalho2022,Hukkanen2023}. It should be noted that in the case of $^{120}$In, the experimental excitation energy of the $5^+$ state derived from the $\beta$ end-point studies was replaced by an extrapolation as the reported values deviate significantly from the mass trends \cite{Wang2021}.

In this work, we report the first Penning-trap measurement of ground-state masses and isomer excitation energies of $^{120-124}$In. Based on the observed production ratios and decay studies, we establish the order of the long-lived states. The results are compared with various theoretical models. 

\section{Experimental method}

The neutron-rich indium isotopes were produced with a 25-MeV proton beam impinging onto a 15 mg cm$^{-2}$ uranium target at the Ion Guide Isotope Separator On-Line (IGISOL) facility at the University of Jyv\"askyl\"a. The fission fragments were stopped in helium gas at a pressure of about 300~mbar, where the charge-state distribution favors singly-charged ions. The ions were extracted from the IGISOL gas cell via a sextupole ion guide \cite{Karvonen2008} to high vacuum, where they were electrostatically accelerated to 30$q$~kV energy ($q$ is the charge of ions). The ions were mass-separated using a 55$^{\circ}$ dipole magnet. The continuous ion beam with the selected mass-to-charge ratio, $A/q$, was then injected into a gas-filled radio-frequency quadrupole \cite{Nieminen2001}, which transformed the continuous beam into ion bunches. Next, the ion bunches were transported to the JYFLTRAP double Penning trap mass spectrometer \cite{Eronen2012}. A post-trap spectroscopy setup was prepared after JYFLTRAP to identify the states of the studied isotopes. The isobarically clean ion bunches purified at JYFLTRAP were implanted in a foil in front of a 500~$\mu$m thick silicon detector, next to which a single GC7020 Ge detector was located.

In the first trap of JYFLTRAP, the ions were cooled, purified and centered using a mass-selective buffer gas cooling technique \cite{Savard1991}. This process was used to select the ions of interest from most isobaric contamination. To isolate the isomeric state from the ground state and from nearby isobars, a Ramsey cleaning method \cite{Eronen2008} was employed for singly-charged $^{120-123}$In ions with Ramsey excitation patterns (On-Off-On) 5-40-5 ms, 5-130-5 ms, 5-15-5 ms and 5-90-5 ms, respectively. The mass measurements of ions with charge-to-mass ratio $q/m$ were performed in the second measurement trap by determining their cyclotron frequency $\nu_c = qB/(2 \pi m)$ in a magnetic field $B$ via a phase-imaging ion-cyclotron-resonance (PI-ICR) technique \cite{Eliseev2013,Eliseev2014,Nesterenko2018}.

The cyclotron frequency measurements of the ions of interest ($\nu_{c}$) were alternated with the similar measurements of the reference ions ($\nu_{c,ref}$), which were interpolated to the time of the actual measurement of the ions of interest, and the cyclotron frequency ratio $r = \nu_{c,ref} / \nu_c$ was determined. The final cyclotron frequency ratio was calculated as a weighted mean of single frequency ratios with the maximum of internal and external uncertainty \cite{Birge1932}.
The atomic mass of the ion of interest is derived from the cyclotron frequency ratio $r$ as
\begin{equation} \label{eq:mass}
M = \frac{z}{z_{ref}} (M_{ref} - z_{ref} m_{e}) r + z m_{e},
\end{equation}
where $M_{ref}$ is an atomic mass of the reference, $m_{e}$ is an electron mass, $z$ and $z_{ref}$ are charge states of the ion of interest and the reference ion, respectively. Only singly- and doubly-charged ions were used in the measurements. The binding energies of the missing electrons were neglected. The ions of $^{133}$Cs with a mass-excess value $\Delta_{lit.} = -88070.943(8)$ keV \cite{Wang2021} were used as a reference for the mass measurements of the studied indium isotopes with an exception of $^{122}$In$^{2+}$ ions for which $^{85}$Rb$^{+}$ (${\Delta_{lit.} = -82167.341(5)}$~keV \cite{Wang2021}) was utilised.

The uncertainty due to the fluctuations of the magnetic field and the uncertainty related to the distortion of the ion motion projection onto the detector, as well the mass-dependent and residual systematic uncertainties in case when the ions of interest and reference ions are not a mass doublet, were taken into account \cite{Nesterenko2021}. When statistically feasible, count-rate class analysis \cite{Roux2013} was performed for determined cyclotron frequency ratios to account for ion-ion interactions in the trap. 

The cyclotron frequency in the PI-ICR method is determined based on the phase difference of the ion's radial motions accumulated in the trap in a phase accumulation time $t_{acc}$. The phase accumulation time was chosen in such a way as to ensure that the ground and isomeric states are separated and that the projection of the cyclotron motion onto the detector is not overlapping with any possible isobaric or molecular contamination. The mass measurements of the Ramsey-cleaned states in $^{120-123}$In$^+$ ions were performed with the phase accumulation time of 759, 545, 1014 and 529~ms, respectively. The phase accumulation time for the measurement with the $^{122}$In$^2+$ ions was 552~ms and for the $^{124}$In$^+$ ions it was 1780~ms to separate the low-lying isomeric state.

\section{Results}

The results of the mass measurements and the comparison with the literature values are summarized in Table~\ref{tab:results}. In the following section, we elaborate on each species.

\begin{table*}
\centering
\caption{\label{tab:results} Ground and isomeric states in $^{120-124}$In studied in this work together with their spin-parities $J^{\pi}$ and half-lives $T_{1/2}$ adopted from the NUBASE2020 evaluation \cite{NUBASE2020}. The frequency ratios $r=\nu_{c,ref}/\nu_{c}$ determined using the PI-ICR technique in this work, corresponding mass-excess values $\Delta$ and excitation energies $E_{x}$ are compared to the literature values ($\Delta_{lit.}$ and $E_{x,lit.}$) from Refs. \cite{Wang2021,NUBASE2020}. The differences between this work and the literature, $\mathrm{Diff.} = \Delta - \Delta_{lit.}$, are added for comparison. The reference nuclides have been listed for each measurement. The values derived from the trends in neighboring nuclei are marked by \#. The states of $^{120x}$In$^+$ and $^{122x}$In$^+$ ions were assigned to a mixture of $5^+$ and $8^-$ near-lying spin states, see text for details. For these two states, an additional 8~keV systematic uncertainty is added when calculating differences with the literature.}
\begin{ruledtabular}
\begin{tabular}{llllllllll}
Nuclide     & $T_{1/2}$ & $J^{\pi}$  & Ref. & $r=\nu_{c,ref}/\nu_{c}$ & $\Delta$ (keV) & $\Delta_{lit.}$ (keV) & $E_x$ (keV) & $E_{x,lit.}$ (keV) & Diff. (keV) \\\hline
$^{120}$In 	& 3.08(8) s & $1^+$\footnotemark[1]		    & $^{120x}$In		        & 0.999 999 193(23)  		& $-85709.1(31)$	& $-85730(40)$  & 	&	& $21(40)$ \\\noalign{\vskip 1.3mm}
\multirow{2}{*}{$^{120x}$In} & 46.2(8) s & $5^+$\footnotemark[1] & \multirow{2}{*}{$^{133}$Cs}	& \multirow{2}{*}{0.902 205 506 (13)} &  \multirow{2}{*}{$-85619.0(16)$\footnotemark[2]}	& $-85680(50)$\#   & \multirow{2}{*}{90.1(26)} & 50(60)\#	& $61(51)$\#\\
 & 47.3(5) s & $8^-$\footnotemark[1]	&   	& 	 &  	& $-85430(200)$\# &  & 300(200)\# 	& $-189(200)$\#\\ \noalign{\vskip 1.3mm}  
$^{121}$In   & 23.1(6) s & $9/2^+$  		& $^{133}$Cs   			& 0.909 727 8574(97) 		& $-85845.0(12)$ 	& $-85835(27)$ & 	&	& $-10(27)$\\
$^{121m}$In  & 3.88(10) m & $1/2^-$  		& $^{121}$In   			& 1.000 002 7875(71) 		& $-85531.1(14)$ 	& $-85521(27)$ & 313.94(80)	&	313.68(7) & $-10(27)$\\ \noalign{\vskip 1.3mm}  
\multirow{2}{*}{$^{122x}$In} 				& 10.3(6) s & $5^+$ \footnotemark[1]		&    \multirow{2}{*}{$^{133}$Cs}			& \multirow{2}{*}{0.917 270 5671(99)} 		& \multirow{2}{*}{$-83550.7(12)$\footnotemark[2]} 	& $-83530(80)$\# & \multirow{2}{*}{0}	& 40(60)\#  & $-21(80)$\#\\
			& 10.8(8) s & $8^-$ \footnotemark[1]		&    			& 		&  	& $-83280(130)$ & 	& 290(140) & $-271(130)$\\ \noalign{\vskip 1.3mm}  
$^{122m}$In & 1.5(3) s & $1^+$ \footnotemark[1] 	    & $^{85}$Rb   			& 0.717 863 0844(82)\footnotemark[3] 	& $-83472.9(13)$ 	&  & 	& &  \\
            &  		    &                    & $^{122x}$In   			& 1.000 000 6694(25) 		& $-83474.6(31)$ 	&  & 	& & \\
            &  		    &                    &    			& Weighted mean: 		& $-83473.45(92)$ 	& $-83571(50)$ & 77.2(15)	& 0 & $98(50)$\\ \noalign{\vskip 1.3mm}              
$^{123}$In  & 6.17(5) s & $9/2^+$  & 	$^{133}$Cs   			& 0.924 795 9723(88) 		& $-83398.6(11)$ 	& $-83429(20)$ & 	&	& $30(20)$\\
$^{123m}$In & 47.4(4) s & $1/2^-$  & 	$^{123}$In   			& 1.000 002 8561(35) 		& $-83071.6(12)$ 	& $-83102(20)$ & 326.99(40)	&	327.21(4) & $30(20)$\\ \noalign{\vskip 1.3mm}  
$^{124}$In  & 3.67(3) s	& $8^-$ \footnotemark[1] &  $^{133}$Cs   			& 0.932 340 247(26) 		& $-80910.5(32)$ 	& $-80890(50)$ & 	&	& $-21(50)$\\
$^{124m}$In & 3.12(9) s	& $3^+$ \footnotemark[1] & $^{124}$In   			& 1.000 000 209(23) 		& $-80886.3(41)$ 	& $-80870(30)$ & 24.2(26)	&	20(60) & $-16(30)$\\
\end{tabular}
\end{ruledtabular}
\footnotetext[1]{The order of the states is based on this work.}
\footnotetext[2]{Consists of only the statistical uncertainty.}
\footnotetext[3]{Measured with doubly-charged $^{122}$In$^{2+}$ ions.}
\end{table*}

\subsection{$^{121,123}$In}

The mass excess of $^{121}$In ($\Delta = -85845.0(12)$ keV) measured in our experiment agrees with the Atomic Mass Evaluation 2020 (AME2020) value ($\Delta_{lit.} = -85835(27)$ keV \cite{Wang2021}) and is 22 times more precise, while for $^{123}$In ($\Delta = -83398.6(11)$ keV) it deviates from the AME2020 value ($\Delta_{lit.} = -83429(20)$ keV \cite{Wang2021}) by $-30(20)$ keV, i.e. by $1.5 \sigma$, and is 18 times more precise. For both odd-A indium isotopes reported in this work, the isomer excitation energy was measured as the mass difference between the isomeric and ground states. Our values of $E_{x,121} = 313.94(80)$~keV and $E_{x,123} = 326.99(40)$~keV for $^{121,123}$In, respectively, are in a good agreement with the precise NUBASE2020 and ENSDF values ($E_{x,121}^{lit.} = 313.68(7)$~keV and $E_{x,123}^{lit.} = 327.21(4)$ keV \cite{ENSDF,NUBASE2020}). 

\subsection{$^{120}$In}

\begin{figure}[htb]
\includegraphics[width=\columnwidth]{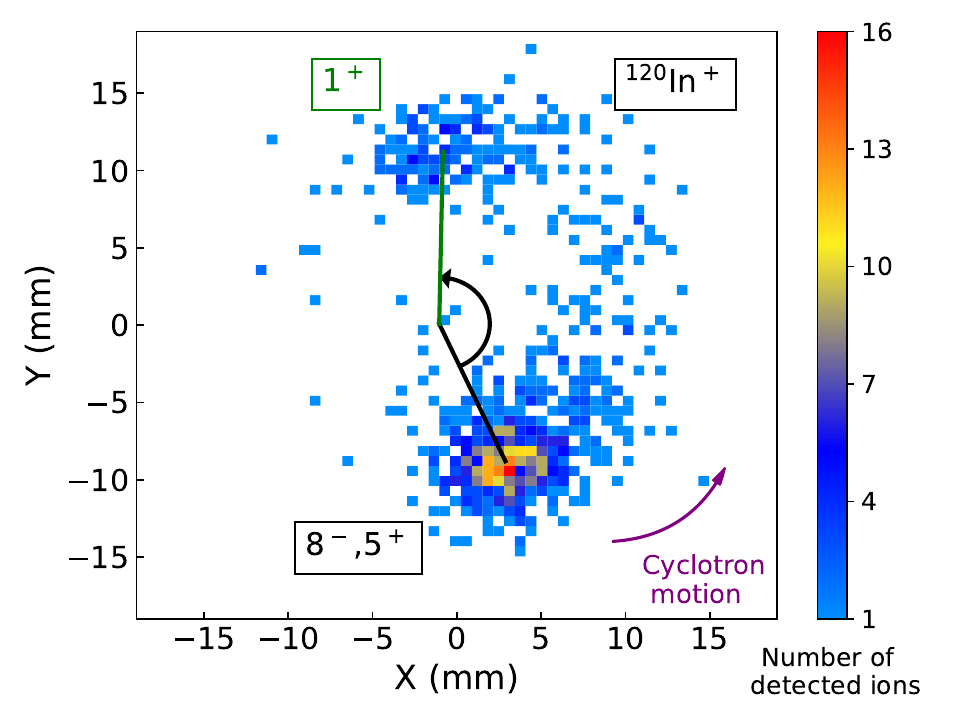}
\caption{\label{fig:in120_PIICR}Projection of the cyclotron motion of $^{120}$In$^+$ ions on the position-sensitive detector obtained with the PI-ICR technique using a phase accumulation time $t_{acc} = 595$~ms. Only bunches with a single detected ion are shown.}
\end{figure}

\begin{figure}[htb]
\includegraphics[width=\columnwidth]{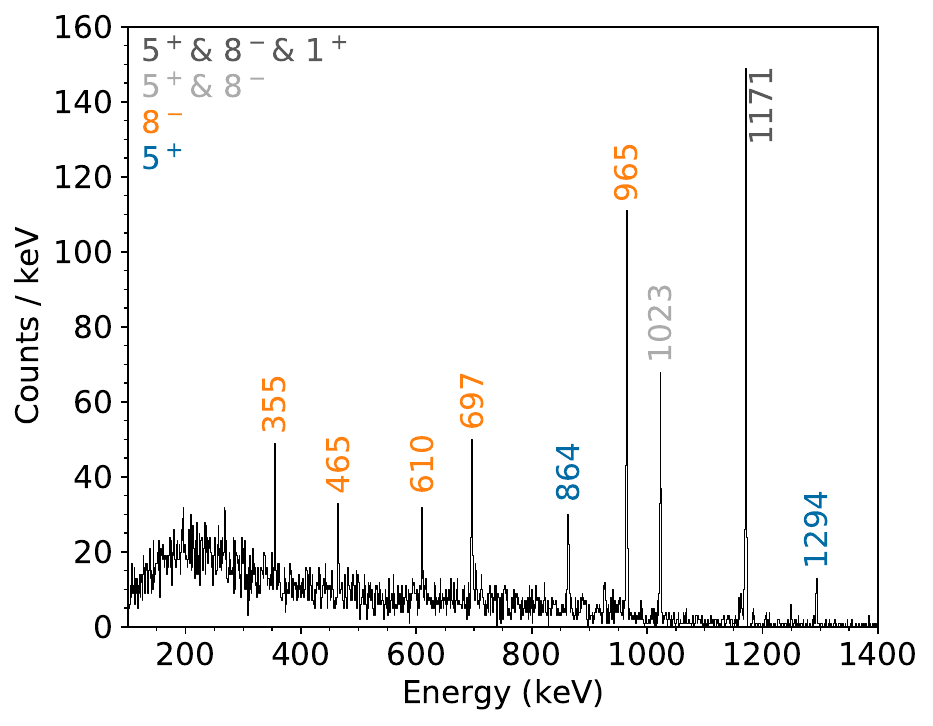}
\caption{\label{fig:in120_spec}$\beta$-gated $\gamma$-ray spectrum obtained for $^{120}$In and $^{120}$Cd produced in fission reactions of $^{nat}$U and purified in the first trap, see text for details. The energies shown in blue and orange indicate the transitions from the $5^+$ and $8^-$ states in $^{120}$In, respectively, in grey the transition that is common for the $5^+$ and $8^-$ states, in black the transition that is common for the $1^+$, $5^+$ and $8^-$ states.}
\end{figure}

Three long-lived states in $^{120}$In are known in the literature \cite{NUBASE2020}. In the PI-ICR measurement, only one state could be clearly observed. However, by limiting the count rate to 1 ion/bunch, a cluster of ions appears at approximately 90 keV below the strongly-produced state (see Fig.~\ref{fig:in120_PIICR}), indicating another state. This behavior was observed with the accumulation times varied between 300 and 856~ms.

The final measurement of the isomeric state was done with an accumulation time of 759~ms, while the excitation energy was determined with accumulation times of 759~ms and 595~ms. The determined ground-state mass excess is  $\Delta = -85709.1(31)$~keV while the isomer excitation energy is $E_{x}=90.1(26)$~keV. The ratio of the strongly-produced isomer to the weakly-produced ground state was about $5.5:1$. 

To further investigate which long-lived states were measured, the isomerically-mixed beam of $^{120}$In was sent to the post-trap decay setup in a continuous mode. The ions of interest were purified by the first trap. However, due to a small frequency difference between $^{120}$In and $^{120}$Cd, the beam contained a small cadmium contamination. The analysis of the $\beta$-gated $\gamma$-ray spectrum ($\Delta T(\gamma - \beta) \leq 250~\mathrm{ns}$) revealed a presence of $\gamma$-ray transitions originating from the decay of the $8^-$ (355, 465, 610, 697 and 965 keV) and $5^+$ (864 and 1294 keV) isomeric states in $^{120}$In (see Fig.~\ref{fig:in120_spec}). Their ratio was determined to be about $3:1$. It should be noted that the presence of the $^{120}$Cd contamination does not influence this observation as $^{120}$Cd decays exclusively to the $1^+$ state in $^{120}$In \cite{ENSDF}. 

Considering that (i) the ratios of the number of ions observed in the PI-ICR measurement ($5.5:1$) and the $\gamma$-ray intensities of the 965- to 864-keV transitions in the decay measurement ($3:1$) differ from each other, (ii) the production of the $5^+$ and $8^-$ states in fission was confirmed and the production of the $1^+$ state cannot be excluded, (iii) production of states with higher spin is favored in fission compared to states with lower spin \cite{Rakopoulos2019} and (iv) an agreement between the ground-state mass-excess from this work (${\Delta = -85708.5(23)}$~keV) and the $1^+$ state mass excess from the AME2020 evaluation (${\Delta = -85730(40)}$~keV \cite{Wang2021}), we assign the $1^+$ state as the ground state. At the same time, the observed PI-ICR spot of the isomeric state is a mixture of the $5^+$ and $8^-$ states. From the spot width, the energy difference between the $5^+$ and $8^-$ states is estimated to be below 15 keV. The uncertainty on the mass-excess value, ${\Delta = -85619.0(16)}$~keV, consists of only the statistical uncertainty and does not take into account the limit on the energy difference between the two long-lived states. 

The $Q_\beta = 5475.7(14)$~keV of the $^{120x}$In isomeric mixture to the $^{120}$Sn ground state extracted in this work is 1.1$\sigma$ away from the AME2020 extrapolation for the $Q_\beta$ value of the first isomeric state ($Q_\beta = 5420(50)\#$~keV \cite{Wang2021}). This estimation replaced two experimental values of $5280(200)$~keV from Ref.~\cite{Kantele1964} and $5340(170)$~keV from Ref.~\cite{Aleklett1978} which were deemed irregular and were proposed as good candidates for new experimental studies \cite{Huang2021}. It should be noted that since the $Q_\beta$ value from this work is based on the mass excess of the isomer mixture, the real $Q_\beta$ of the first isomeric state is lower and, consequently, closer to the AME2020 estimate.

\subsection{$^{122}$In}

\begin{figure}[htb]
\includegraphics[width=\columnwidth]{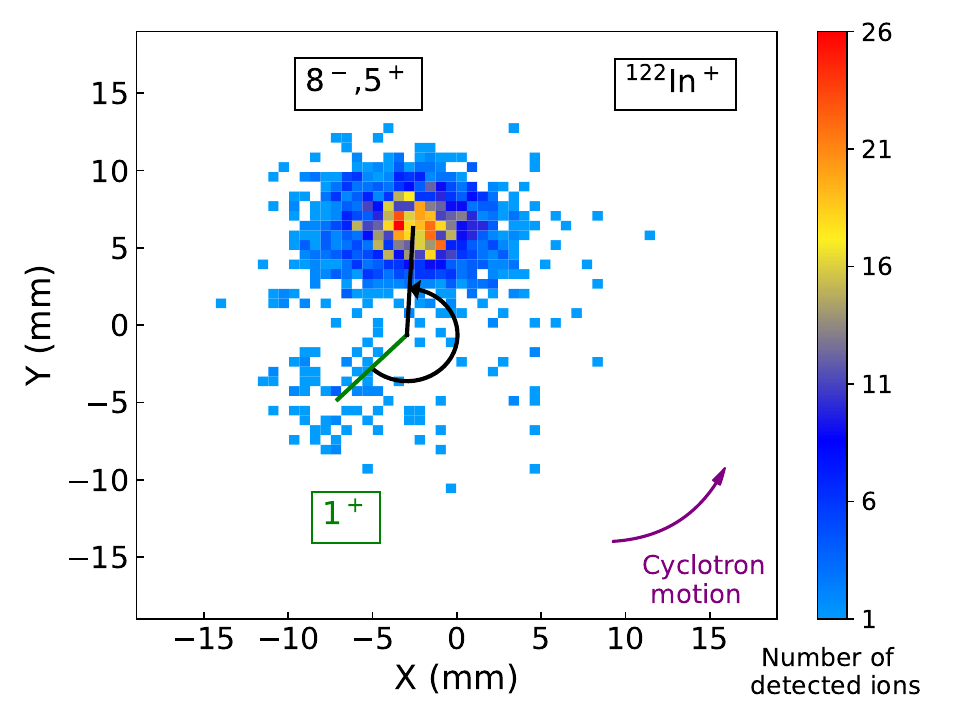}
\caption{\label{fig:in122_PIICR}Projection of the cyclotron motion of $^{122}$In$^+$ ions on the position-sensitive detector obtained with the PI-ICR technique using a phase accumulation time $t_{acc} = 1014$~ms. Ions of $^{122}$In were produced in proton-induced fission of $^{nat}$U. The spin assignment is based on the post-trap spectroscopy measurements and the mass measurement of $^{122}$In$^{2+}$ ions produced after $\beta$-decay of $^{122}$Cd, see text for details.}
\end{figure}

Three long-lived $\beta^-$-decaying states in $^{122}$In are known in the literature \cite{NUBASE2020}. To determine their masses and order, two measurements were performed: the first one with the $^{122}$In$^{+}$ ions produced directly in fission, while the second one with the $2^+$ ions produced via in-trap decay of $^{122}$Cd$^+$.  

In the first measurement, only two spots (states) were observed despite variation of the phase accumulation time $t_{acc}$ between 363~ms and 1300~ms. The final measurement was performed with $t_{acc} = 1014$~ms (see Fig.~\ref{fig:in122_PIICR}) resulting in the mass-excess values of $\Delta = -83550.7(12)$~keV and $\Delta = -8383474.6(31)$~keV for the ground-state and the isomer, respectively. The ratio of number of ions between lighter state and heavier state was about $20:1$. 

During the second measurement, the fission-produced ions of $A/q = 122$, containing also $^{122}$Cd$^+$ were captured in the first trap and stored there for 500~ms to allow $\beta$ decay to take place. The $^{122}$In$^{2+}$ ions were purified by using a buffer-gas technique \cite{Savard1991} and sent to the second trap for the mass measurements. In the PI-ICR measurement, only one state was observed and its mass was measured with $t_{acc} = 552$~ms using $^{85}$Rb$^+$ as reference ions. The measured mass excess (${\Delta = -83472.9(13)}$~keV) coincides with the mass excess of the heavier state of $^{122}$In produced directly in fission (${\Delta = -83474.6(31)}$~keV). 

To further investigate which states in $^{122}$In were observed, the purified beams of $^{122}$Cd and $^{122}$In were sent to the post-trap decay spectroscopy setup. In the case of the $^{122}$Cd beam, the ions produced in uranium fission were purified in the first trap and collected using the post-trap spectroscopy setup with a cycle of 20 s implantation and 20 s decay time. Only a weak 1140-keV $\gamma$-line originating from the $\beta$-decay of $^{122}$In was observed in the $\beta$-gated $\gamma$-ray spectrum which is consistent with an exclusive production of the $1^+$ state in $^{122}$In in the $\beta$-decay of $^{122}$Cd \cite{ENSDF}. This observation allows us to unambiguously assign the $1^+$ spin-parity to the observed isomeric state in $^{122}$In.

\begin{figure}[htb]
\includegraphics[width=\columnwidth]{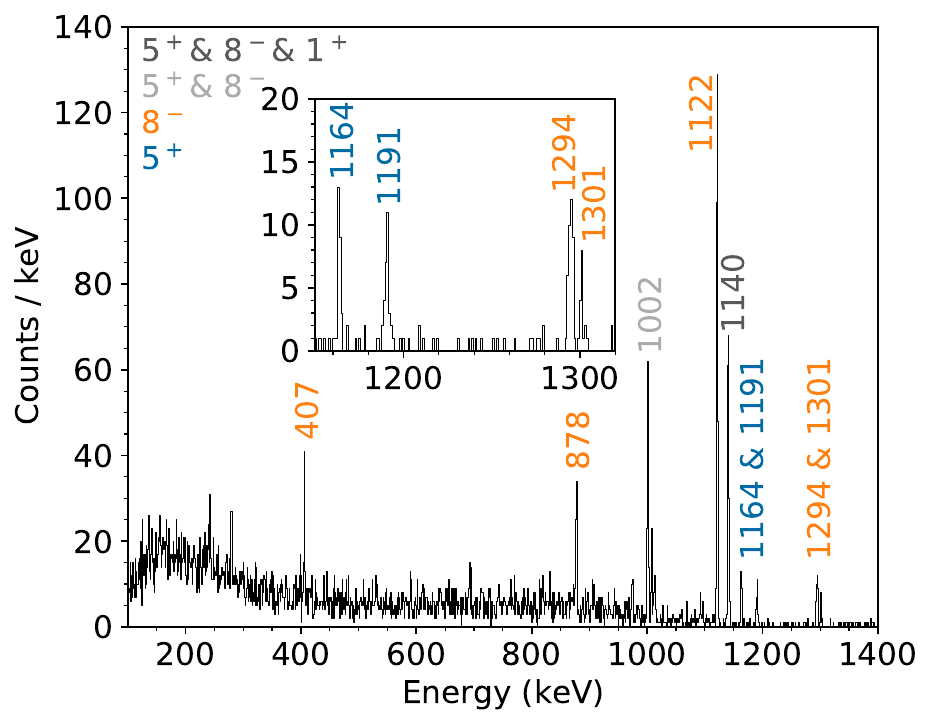}
\caption{\label{fig:in122_spec}$\beta$-gated $\gamma$-ray spectrum obtained for $^{122}$In produced in fission reactions of $^{nat}$U and purified in the first trap. The energies shown in blue and orange indicate the transitions from the $5^+$ and $8^-$ states in $^{122}$In, respectively, in grey the transition that is common for the $5^+$ and $8^-$ states, in black the transition that is common for the $1^+$, $5^+$ and $8^-$ states.}
\end{figure} 

In the case of the $^{122}$In beam, the ions were sent to the spectroscopy setup with a cycle of 30 s implantation and 30 s decay time. In the prompt $\beta$-gated $\gamma$-ray spectrum, $\gamma$-ray transitions assigned to the decay of the $8^-$ state (407, 878, 1122, 1294 and 1301 keV) and the $5^+$ state (1164 and 1191 keV) were observed \cite{Fogelberg1979} (see Fig.~\ref{fig:in122_spec}). The yield ratio between the $8^-$ and $5^+$ states, based on the 1122- and 1164-keV $\gamma$-rays, is about $3:1$. Considering that (i) only two spots were clearly visible in the PI-ICR measurement of the fission-produced $^{122}$In, (ii) the number-of-ions ratio between lighter and heavier states ($20:1$) is not consistent with the $8^-$-to-$5^+$ ratio from the spectroscopy studies ($3:1$), (iii) the heavier observed state is assigned the $1^+$ spin-parity and (iv) the $\beta$-particles' decay curve of the $^{122}$In beam is consistent with a small admixture of the $1^+$ state, one can conclude that the large phase spot in the PI-ICR measurement corresponds to the unresolved $8^-$ and $5^+$ states ($\le$15~keV), see Fig.~\ref{fig:in122_PIICR}. 

Compared to NUBASE2020, we change the order of the long-lived states in $^{122}$In with the $5^+$ or $8^-$ states being the ground state and the $1^+$ state being the second isomer. However, the NUBASE2020 ordering was based on extrapolations and $\beta$ end-point studies \cite{NUBASE2020}.

The averaged mass-excess value for the ${8^- + 5^+}$ states is ${\Delta = -83550.7(12)}$~keV. This value is in agreement with the AME2020 ground-state value (${\Delta_{lit.} = -83571(50)}$~keV \cite{Wang2021}) which was previously assigned to the $1^+$ state. However, it should be noted that the authors of the original measurement \cite{AjzenbergSelove1978} did not assign the ground state to any long-lived states in $^{122}$In and it was done by the evaluators based on other available data. As in the case of $^{120}$In, the uncertainty does not take into account the fact it is a mixture of two long-lived state. The weighted mass-excess value of two measurements for the $1^+$ isomer is ${\Delta = -83473.45(92)}$~keV and the energy difference between $1^+$ and $8^- + 5^+$ states is 77.2(15)~keV. 

In addition to the $\gamma$ spectroscopy, the decay cycle used for the cadmium beam enabled extraction of the $^{122}$Cd half-life. The Bateman's equations were fitted to the $\beta$-particles' decay curve collected with the silicon detector and the $^{122}$In($1^+$) state half-life ($T_{1/2}$ = 1.5(3)~s \cite{NUBASE2020}) was provided as a prior. Our result, $T_{1/2}$ = 5.98(10)~s, is significantly longer than 5.24(3)~s from Ref. \cite{Rudstam1981}, 3.13(12)~s from Ref. \cite{Grapengiesser1974} and 5.5(1)~s from Ref. \cite{Grappengiesser1970} but it is in a very good agreement with 5.91(12)~s \cite{Grapengiesser1974a} and 5.78(9)~s \cite{Scheidemann1973}.

\subsection{$^{124}$In}

Due to large uncertainties, the order of the ground and isomeric states of $^{124}$In in NUBASE2020 is not well established \cite{NUBASE2020}. In particular, the $3^+$ state (${\Delta_{lit.} = -80870(30)}$~keV) is proposed to be the ground state despite the $8^-$ state being lighter (${\Delta_{lit.} = -80890(50)}$~keV). The $^{124}$In experimental mass-excess value ${\Delta = -80910.5(32)}$~keV from our work agrees with the AME2020 value for the $8^-$ state and it is 16 times more precise. The excitation energy of the studied isomeric state $^{124m}$In is $E_{x} = 24.2(26)$~keV and it is in agreement with the NUBASE2020 value (${E_{x,lit}=20(60)}$~keV \cite{NUBASE2020}). 

The ground-to-isomeric-state production ratio in the PI-ICR measurement was about $2.8:1$. Based on the fact that the states with higher spins are typically more populated in proton-induced fission of $^{nat}$U \cite{Rakopoulos2019}, considering both states have similar half-lives and taking into account the agreement between the NUBASE2020 and this work mass-excess values, we assign the $8^-$ spin-parity to the ground state and $3^+$ to the isomer.

Based on our results, we can revise the $^{124}$Cd decay scheme reported in Ref. \cite{Batchelder2016}. By shifting all the levels by the excitation energy of the $3^+$ state (24.2(26)~keV), we note that the $1^+_1$ state is now located at 122.9(26)~keV and it overlaps with the state at 122(15)~keV which was previously observed in the (t,$^{3}$He) reaction studies \cite{AjzenbergSelove1978}.

\section{Discussion}

The experimental masses reported in this work were compared with theoretical predictions of the Skyrme nuclear Density Functional Theory (DFT) model calculated using four Skyrme interactions: SLy4 \cite{Chabanat1998}, SV-min \cite{Klupfel2009}, UNEDF0 \cite{Kortelainen2010} and UNEDF1 \cite{Kortelainen2012}, as well as two Skyrme-Hartree-Fock-Bogoliubov (Skyrme-HBF) models, BSkG1 \cite{Scamps2021} and BSkG2 \cite{Ryssens2022}. In addition, we provide shell-model calculations for $^{124}$In and we compare it with known experimental results.

\begin{figure}[h!t!b]
\includegraphics[width=\columnwidth]{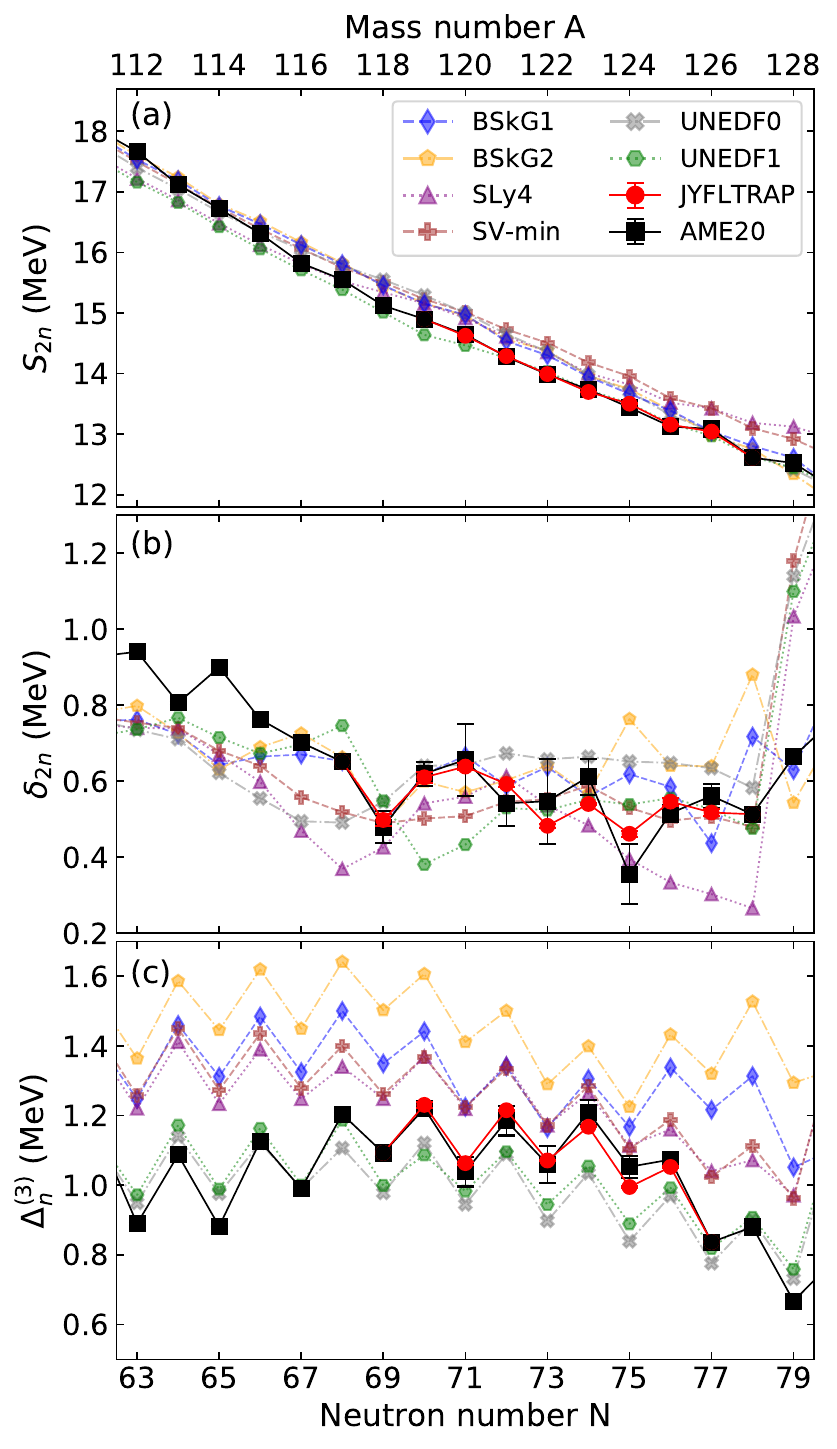}
\caption{\label{fig:theoryshort}A comparison of two-neutron separation energies $S_{2n}$ (panel a), two-neutron shell-gap energies $\delta_{2n}$ (panel b) and three-point neutron gaps $\Delta^{(3)}_n$ (panel c) between experimental values from this work and AME2020 \cite{Wang2021}, and the theoretical models: BSkG1 (blue diamonds), BSkG2 (orange pentagons), SLy4 (purple triangles), SV-min (brown $+$ symbols), UNEDF0 (grey $x$ symbols) and UNEDF1 (green hexagons). For the ground state of $^{122}$In, only the statistical uncertainty was included.}
\end{figure}

First, we calculated two-neutron separation energy $S_{2n}$ defined as
\begin{equation}
S_{2n}(Z,N) = \Delta(Z,N-2) - \Delta(Z,N) +2\Delta_n \mathrm{,}
\end{equation}
where $\Delta(Z,N)$ is a mass excess of a nucleus with a given $Z$ and $N$ and $\Delta_n$ is a mass excess of a free neutron. $S_{2n}$ is a sensitive probe to study structural changes, for instance (sub)shell closures, in isotopic chains. A comparison between experimental results from this work and AME2020 \cite{Wang2021} as well as the theoretical models is presented in Fig.~\ref{fig:theoryshort}a. 

Overall, there is a good agreement between this work and AME2020; however, our values have much smaller uncertainties. The $S_{2n}$ values in the region reported in this work ($120 \leq A \leq 126$) is reproduced only by the DFT model with the UNEDF1 interaction. Other models are systematically overestimating this observable by about 400 keV. However, it should be noted that around $A=115$ there is a considerable shift in the $S_{2n}$ trend which is not reproduced by any model.

To further study the evolution of $S_{2n}$, the two-neutron shell gaps $\delta_{2n}$:
\begin{equation}
\begin{split}
\delta_{2n}(Z,N) &= S_{2n}(Z,N) - S_{2n}(Z,N+2) =\\
&= \Delta(Z,N+2) + \Delta(Z,N-2) - 2\Delta(Z,N)\mathrm{,}
\end{split}
\end{equation}
are also calculated and presented in Fig.~\ref{fig:theoryshort}b. Compared to the AME2020 values, the trend is flattened, in particular at $A=124$. This indicates that the slope of the $S_{2n}$ curve stays constant. At the same time, none of the presented models is able to reproduce the trend of this observable in the region reported in this work. In addition, all the interactions used with the DFT model significantly overestimate the $\delta_{2n}$ value at $A=128$ and this behavior is not repeated by the Skyrme-HBF models. 

The odd-even staggering parameter $\Delta_{n}^{(3)}$, which is understood as the energy gap between masses of odd- and even-$N$ isotopes, is often associated with the neutron pairing gap. In addition, it can be also used as a probe for deformation change or subshell close in the isotopic chain \cite{Bender2000,Duguet2001}. It is defined as
\begin{equation}
\begin{split}
\Delta_{n}^{(3)}(Z,N) =& \frac{(-1)^{N}}{2} \big[\Delta(Z,N+1)\\&+\Delta(Z,N-1)-2\Delta(Z,N)\big] \mathrm{.}
\end{split}
\end{equation}
and it is plotted in Fig. \ref{fig:theoryshort}c. 

The analysis of the theoretical $\Delta_{n}^{(3)}$ values for the indium isotopic chain shows that they are split in two groups. The first group consists of the DFT calculations with the UNEDF$^*$ family interaction while in the second one there are calculations from the Skyrme-EDF and the remaining DFT-based models. While the models from the first group are reproducing the experimental data and the $\Delta_{n}^{(3)}$ trends rather well, in the region reported in this work they systematically underestimate the reported values. On the other hand, the models from the second group systematically overestimate $\Delta_{n}^{(3)}$ and they are the closest to the experimental values around $N=73$. 

\begin{figure}[h!t!b]
\includegraphics[width=\columnwidth]{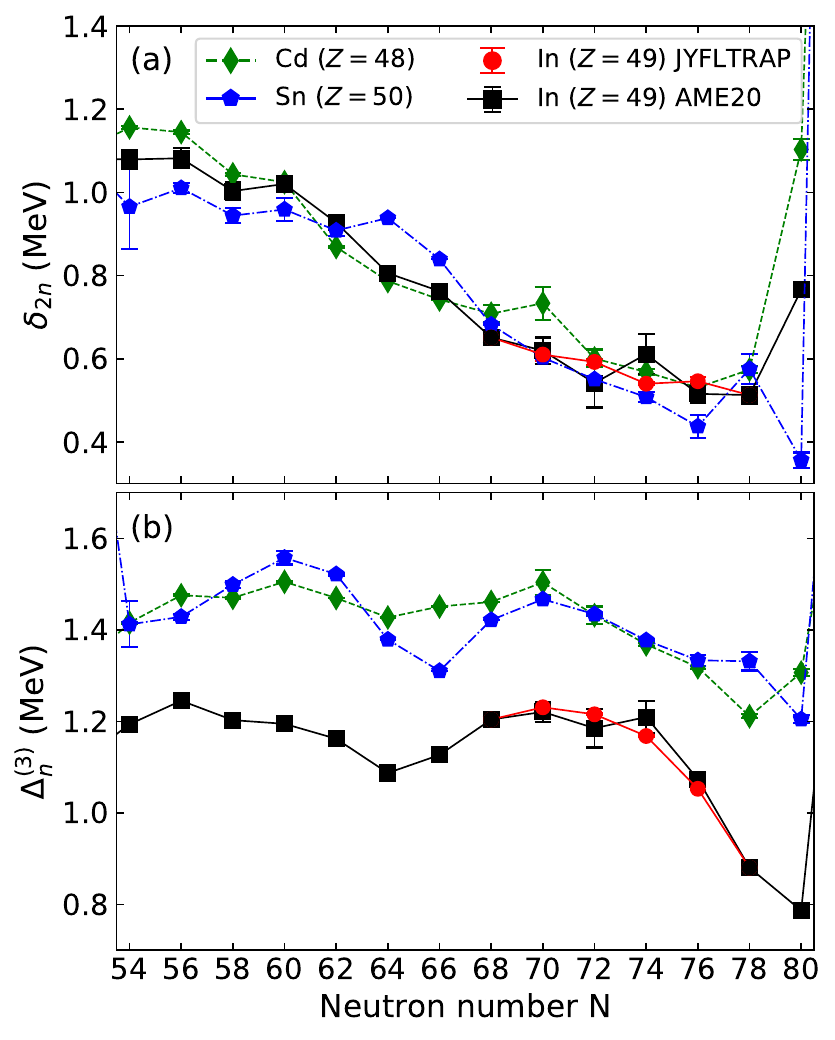}
\caption{\label{fig:trensdCdInSn}A comparison of a) two-neutron shell-gap energies $\delta_{2n}$ and b) three-point neutron gaps $\Delta^{(3)}_n$ for the even-$N$ cadmium, indium and tin isotopic chains. Data is taken from AME2020 \cite{Wang2021} and this work. For the ground state of $^{122}$In, only the statistical uncertainty was included.}
\end{figure}

The comparison of the $\delta_{2n}$ and $\Delta_{n}^{(3)}$ values for the even-$N$ species between the neighboring isotopic chains of indium ($Z=49$), cadmium ($Z=48$) and tin ($Z=50$) is presented in Fig. \ref{fig:trensdCdInSn}. For the $\delta_{2n}$ values one can notice that the trend for the indium and cadmium isotopic chains is almost identical, especially after adding the new experimental masses reported in this work. For all three elements a shift around $N=66$ can be observed; however, it is more abrupt for the tin chain. On the other hand, for the $\Delta_{n}^{(3)}$ values differences are more pronounced. In tin isotopes, there is a well-defined minimum at $N=66$, while in indium it is shifted towards $N=64$ and it is more shallow. For the cadmium chain this feature is not present. 

In Ref.~\cite{Togashi2018}, the discontinuity in $\delta_{2n}$ in tin isotopes was linked to the significant breaking of the $Z=50$ magic number and interpreted as a quantum phase transition. A similarity in the behavior of this observable in the indium isotopic chain might indicate that this phenomenon may be also present. 

\begin{figure*}
\includegraphics[width=0.9\textwidth]{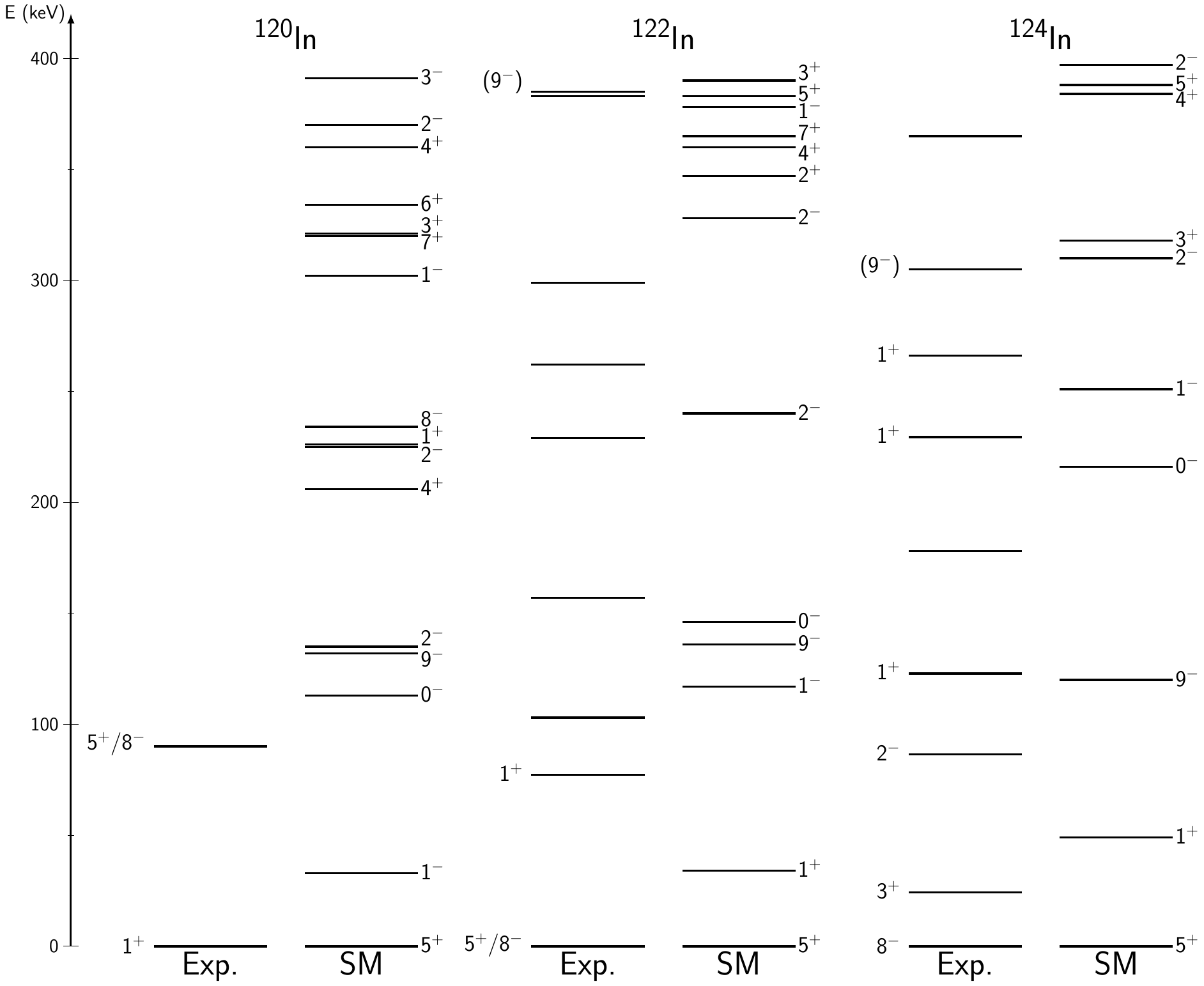}
\caption{\label{fig:InSM}Comparison of the experimental (Exp.) and shell model (SM) schemes of excited states in $^{120,122,124}$In up to 400 keV. The experimental scheme is based on this work and adjusted results from Refs. \cite{AjzenbergSelove1978,Rejmund2016,Batchelder2016}.}
\end{figure*}

The shell model calculations were performed for the odd-odd $^{120,122,124}$In species with the jj45pna interaction \cite{Machleidt2001} using the KSHELL code \cite{Shimizu2013,Shimizu2019}. While a quantitative comparison with the theory is hindered by a limited amount of experimental information (see Fig. \ref{fig:InSM}), a few general observations can be drawn. 

In all three cases the ground states are predicted to be $5^+$. Based on this work it might be correct for $^{122}$In; however, for $^{120,124}$In our results unambiguously excluded that possibility. Also, the $9^-$ states are predicted to be lower in energy that the $8^-$ states which is not in agreement with the laser spectroscopy results \cite{Vernon2019}. It should be noted that in the case of $^{124}$In the number of calculated low-lying $1^+$ states is too low compared to the $\beta$-decay study \cite{Batchelder2016}. Also, the predicted number of isomers and their spins-parities are not in agreement with the experimental observations. 

Some of the aforementioned discrepancies might be explained by an absence of proton excitation across the magic $Z=50$ shell. While the jj45pn interaction has a limited valence space and does not allow for particle excitation across $Z=50$ or $N=82$, it has been successfully used to explain the structure of excited states in the isotopes lying the vicinity of $^{100,132}$Sn doubly-magic nuclei \cite{Lorenz2019,Nesterenko2018,Jin2021,Wang2021a,Wang2022}. At the same time, it has been previously observed that an increase of valence space and excitation across magic numbers were necessary to explain experimental phenomena observed around the nickel ($Z=28$) \cite{Leoni2017,Olaizola2017,Stryjczyk2018,Yang2018,Olaizola2019,Ichikawa2019} and the lead ($Z=82$) \cite{Marsh2018,Sels2019} neutron mid-shells regions. However, further theoretical studies on indium isotopes are needed for a better understanding of the measured species.

\section{Conclusion}

The mass measurements of the ground and isomeric states in $^{120-124}$In isotopes have been performed at the JYFLTRAP double Penning trap mass spectrometer using the PI-ICR technique. The directly measured mass values of the ground states were significantly improved compared to the mass values derived in AME2020 \cite{Wang2021}. The excitation energies of the isomers in $^{121m,123m}$In are in a good agreement with the precisely-known values in NUBASE2020 \cite{NUBASE2020} while in $^{124m}$In it was determined for the first time. In $^{120,122}$In isotopes, it was possible to separate the $1^+$ state and the energy difference between the $5^+$ and $8^-$ states was determined to be $\le$15~keV. Based on the mass measurement of the in-trap decay of $^{122}$Cd ions, the second isomeric state in $^{122}$In was unambiguously assigned as the $1^+$ state. Presence of three known long-lived states in $^{120,122}$In was confirmed by combining decay spectroscopy results and ratios of the numbers of detected ions. Based on this ratio, the ground state of $^{124}$In was assigned to be the $8^-$ state and the isomer to be the $3^+$ state. The comparison of the experimental data with different theoretical models revealed systematic problems in a description of the indium isotopic chain. 

\begin{acknowledgments}

Funding from the European Union’s Horizon 2020 research and innovation programme under grant agreements No 771036 (ERC CoG MAIDEN) and No 861198–LISA–H2020-MSCA-ITN-2019 are gratefully acknowledged. J.Ru. acknowledges financial support from the Vilho, Yrj\"o and Kalle V\"ais\"al\"a Foundation. M.H. acknowledges financial support from the Ellen \& Artturi Nyyss\"onen foundation. T.E. and A.d.R. acknowledge support from the Academy of Finland project No. 295207, 306980 and 327629. D.Ku. acknowledges the support from DAAD grant number 57610603.

\end{acknowledgments}

\bibliographystyle{apsrev}
\bibliography{mybibfile}

\end{document}